\documentclass[twocolumn,prb,amsmath,amssymb,showpacs]{revtex4}

\usepackage{bm}

\ifx\pdftexversion\undefined
  \usepackage[dvips]{graphicx}
\else
  \usepackage[pdftex]{graphicx}
\fi

\def \be{\begin{equation}}
\def \ee{\end{equation}}
\def \bmlett{\begin{mathletters}}
\def \emlett{\end{mathletters}}

\def \NN{{\mathcal N}}

\begin{document}



\title{Laser-like Instabilities in Quantum Nano-electromechanical Systems}

\author{S. D. Bennett and A. A. Clerk}
\affiliation{ Department of Physics, McGill University, Montr\'{e}al,
 Qu\'{e}bec, Canada, H3A 2T8}
\date{Sept. 18, 2006}

\begin{abstract}
We discuss negative damping regimes in quantum nano-electromechanical
systems formed
by coupling a mechanical oscillator to a single-electron
transistor (normal or superconducting).  Using an analogy to a
laser with a tunable atom-field coupling, we demonstrate how these
effects scale with system parameters.  We also discuss the
fluctuation physics of both the oscillator and the single-electron
transistor in this regime, and the degree to which the
oscillator's motion is coherent.
\end{abstract}
\maketitle

At their heart, quantum nano-electromechanical systems (NEMS)
consist of nothing more than a mechanical oscillator coupled to a
mesoscopic conductor.  Despite their apparent simplicity, they
have been the focus of considerable recent interest.  It has been
predicted and shown experimentally that a single-electron
transistor (SET) coupled to a mechanical oscillator may be used
for near quantum-limited position detection \cite{Blencowe00,
Knobel03,LaHaye04,Naik06,Blencowe04}.  An equally fascinating
aspect of these systems is their underlying quantum dissipative
physics: from the point of view of the oscillator, the mesoscopic
conductor acts as a non-equilibrium bath which both heats and
damps the oscillator \cite{Mozyrsky02,Armour04,Clerk04c}.  For a
sufficiently weak oscillator-conductor coupling, and a
sufficiently high-Q oscillator,  one can map the conductor onto an
effective equilibrium bath.  Even in this regime, novel effects
may take place. Recent work shows that in systems with either
normal-metal or superconducting SET's, one can operate in regimes
where the SET generates negative damping of the
oscillator\cite{Blanter05both,Blencowe05,Clerk05}.  The resulting
instability is ultimately cut off by a nonlinearity in the
dynamics, and is characterized by an effective strong coupling
between the mechanical and electronic dynamics.

In this paper, we investigate the properties of both normal-state
and superconducting SET NEMS systems in the negative damping
regime.  Using the analogy between these systems and a laser with
a tunable atom-field coupling, we derive simple scaling relations
for the stationary state and fluctuations in the negative damping
regime.  We also discuss how noise measurements could be used to
sensitively probe the physics in this regime.  In particular, the
critical slowing down associated with the transition to the
``lasing" state can be seen clearly in the low-frequency current
noise of the transistor.  Note that the analogy between laser
physics and a superconducting SET NEMS was also studied recently
in Ref. \onlinecite{Armour06} using a numerical approach. An
alternate proposed mechanical analogue of a laser was discussed in
Ref. \onlinecite{Bargatin03}.

{\it Model}. We consider a standard SET-based NEMS where the
oscillator acts as a voltage gate with an $x$-dependent coupling
capacitance to the SET island; details may be found in
Ref.~\onlinecite{Blencowe04}. In the absence of any intrinsic
damping of the oscillator, the total Hamiltonian is given by $H =
H_{osc} + H_{0} + H_{C}$.  Here, $H_{osc}$ describes an oscillator
with angular frequency $\Omega$ and mass $m$.  $H_{0}$ describes
the non-interacting part of the SET Hamiltonian, and includes the
kinetic energies of electrons in the leads and island, as well as
tunnelling terms taking electrons to and from the leads; in the
case of a superconducting SET (SSET), the island and leads are
described by BCS Hamiltonians.  Finally, $H_C$ describes the
Coulomb charging energy of the island in the presence of the
oscillator:
\begin{equation}
    H_{C} = E_C \left[ \hat{n}^2 - 2 \hat{n} \left(
     \NN_0 + \frac{A}{2 E_C} \hat{x} \right) \right],
\end{equation}
where $E_C$ is the charging energy of the SET island, $\hat{n}$ is
the island charge, and $\NN_0$ is the dimensionless gate voltage
associated with a fixed control gate. The coupling strength is
given by $A / 2E_C = (V_{osc}/e) \cdot dC_{osc}/dx$, where
$V_{osc}$ is the voltage bias on the oscillator and $C_{osc}$ is
the capacitance between the oscillator and the SET
island\cite{Clerk05}. In what follows, the oscillator is always in
the regime where $C_{osc}$ depends linearly on $x$. We see that
$\hat{n}$ acts as a back-action force on the oscillator, while the
SET experiences an effective $x$-dependent gate voltage $\NN[x] =
\NN_0 + (A / 2E_C) \cdot x$.  In the usual case where $\Omega$ is
much smaller than the typical tunnelling rates in the SET, and the
coupling strength $A$ is sufficiently weak, one can combine
linear-response theory with a Born-Oppenheimer approximation to
rigorously derive a classical Langevin equation describing the
oscillator \cite{Clerk05}:
\begin{equation}
    m \ddot{x}  =  -(k + \Delta k[x]) x -
        m \left(\gamma_0 + \gamma[x] \right) \dot{x} + \delta f_0
    + \delta f[x].
    \label{eq:Langevin}
\end{equation}
The back-action force leads to a damping term $\gamma[x]$,
spring-constant renormalization $\Delta k[x]$, and a stochastic
force $\delta f(t)$ with a white, $x$-dependent spectral density
$S_{\delta f}[x]$. The damping and noise are determined by the
quantum noise spectrum $S_n(\omega,\NN) \equiv \int dt e^{i \omega
t} \langle \hat{n}(t) \hat{n}(0) \rangle_{\NN}$ of the SET island
charge fluctuations, evaluated at zero coupling to the oscillator
and for a fixed gate voltage $\NN$; $\Delta k$ is determined by
the corresponding charge susceptibility:
\begin{align}
    m\gamma[x] & =
        \frac{A^{2}}{\hbar}
        \frac{dS_n(\omega,\NN)}{d \omega}\bigg|_{\omega = 0,\ \NN = \NN[x]},
    \label{eq:LangDamping}  \\
    S_{\delta f}[x] & =
            A^2 S_n \big(\omega=0,\NN[x] \big),
    \label{eq:LangNoise}    \\
    \Delta k[x] & =
        \frac{A^2}{2E_C}
        \frac{d \langle n \rangle}{d \NN_0}\bigg|_{\NN = \NN[x]}.
    \label{eq:Deltak}
\end{align}
The spectrum $S_n(\omega,\NN)$ can been calculated for various SET
processes using standard techniques \cite{Clerk05}.
Physically, Eqs. (\ref{eq:LangDamping}--\ref{eq:Deltak}) are based
on the approximation that at each instant in time, the oscillator
sees the SET as an effective bath whose properties are determined
by the instantaneous gate voltage $\NN[x]$. The separation of
timescales required for this description to be valid is well
satisfied in current experiments \cite{Knobel03,LaHaye04,Naik06}.
Eq.~(\ref{eq:Langevin})  also includes damping and noise terms due
to coupling to an equilibrium bath: in the absence of the SET, the
intrinsic quality factor of the oscillator is $Q_0 = \Omega /
\gamma_0$ and its temperature $T_0$ is determined by the strength of the
force $\delta f_0(t)$.

{\it Negative Damping}. As the SET is voltage biased and hence out
of equilibrium, it is possible for the back-action damping
$\gamma$ to become negative and even overwhelm the positive bath
damping $\gamma_0$.  For a SSET, negative damping can arise near
operating points corresponding to incoherent Cooper-pair tunneling
in the single/double Josephson quasiparticle processes (JQP/DJQP)
\cite{Clerk05,Blencowe05}.  If SET voltages are chosen so that
tunneling Cooper-pairs can move closer to resonance by emitting
energy to the oscillator, one generically gets negative damping.
Negative damping can also arise in a normal metal SET if one has
strongly energy-dependent tunnel matrix elements and a resulting
non-monotonic $\langle n \rangle$ versus $\NN$ curve
\cite{Blanter05both, Clerk05}.

For either of the above two cases, negative damping means that the
stationary state of the oscillator will be determined by the
nonlinearity of the SET-induced damping (i.e. the fact that
$\gamma$ is a function of $x$).  Crudely speaking, the stationary
state corresponds to an oscillator amplitude large enough that the
oscillating gate voltage $\NN[x]$ smears out the negative damping
from the SET, making the total damping zero.  A convenient way to
describe this physics is to make use of the high quality factors
of typical NEMS.  One can then derive a Kramers-like equation for
the energy distribution function $w(E,t)$ of the oscillator
\cite{Blanter05both, Clerk05}:
\begin{eqnarray}
    \frac{d}{dt} w = \frac{\partial}{\partial E} E \left[
        \gamma_0 + \gamma(E) + \frac{D_0 + D(E)}{m} \frac{\partial}{\partial E}
         \right] w,
    \label{eq:FP}
\end{eqnarray}
where the energy-dependent damping and diffusion are given by
\begin{eqnarray}
    \gamma(E) & = & 2 \int_0^{2 \pi} \frac{d \theta}{2 \pi}
        \gamma [x]  \cos^2 \theta,
            \label{eq:GammaE}
        \\
    D(E) & = &  \int_0^{2 \pi} \frac{d \theta}{2 \pi}
        S_{\delta f} [x] \cos^2 \theta,
\end{eqnarray}
with $x=\sqrt{2E/k}\sin{\theta}$. The stationary solution of Eq.
(\ref{eq:FP}) is given by a generalized Boltzmann distribution.
Defining $\widetilde{T}_{osc}(E) \equiv \left[D(E) + D_0\right] /
[\gamma(E) + \gamma_0]$, we have
\begin{eqnarray}
    w(E) \propto \exp \left[
        - \int_0^{E} \frac{d E'}{k_B \widetilde{T}_{osc}(E')}
        \right].
    \label{eq:pEstat}
\end{eqnarray}

In the negative damping regime of interest, $\gamma(E) + \gamma_0$
is negative at $E=0$, and becomes positive at large enough $E$. In
the simplest case, $\gamma(E)$ is simply a monotonic increasing
function of E; this arises in a SSET if one detunes the
Cooper-pair resonances to achieve a maximum negative damping
\cite{Clerk05}.  One then finds that $w(E)$ has a maximum at
$E=E_0$, where $E_0$ is defined by $\gamma(E_0) + \gamma_0 = 0$.
Moreover, if $E_0$ is sufficiently large, the distribution has a
Gaussian form, with a width given by
\begin{eqnarray}
    \sigma^2 & = &
        \frac{D_0 + D(E_0)}{m}
        \left[\frac{d \gamma}{d E} \Big|_{E=E_0} \right]^{-1}
    \label{eq:sigma}.
\end{eqnarray}

The negative damping instability discussed here in the context of
a quantum NEMS is reminiscent of the physics of a laser.  In the
typical case of a single mode cavity laser with fast atomic
relaxation, the effects of the population-inverted atoms on the
relevant cavity mode can be described in terms of
amplitude-dependent damping that is negative for small mode
occupancies \cite{Haken75}.  In our case, it is instead tunnelling
electrons or Cooper pairs in the SET which provide the negative
damping.  Note also that our 
system is also very similar to a non-resonantly driven oscillator coupled non-linearly to 
an oscillator bath; the negative damping regime here was studied
extensively (both classically and quantum mechanically)
by Dykman and Krivoglaz.\cite{Dykman84}  In what follows, we explore in more detail the extent to
which our system is analogous to a laser.  

{\it Scaling}. Two basic features of a laser are both a large
average number of quanta in the lasing mode, and relatively small
number fluctuations described by Poisson statistics.  We would
thus like to know how these quantities scale with system
parameters in the SET-based NEMS considered here.  We first assume
that the SET gate voltage $\NN_0$ has been set to maximize the
negative damping.  We then introduce the characteristic scale for
the gate voltage, $\NN^*$, which is defined as the change of the
gate voltage required to significantly reduce the magnitude of the
SET damping \footnote{The precise definition chosen for the scale
$\NN^*$ is not important. Changing $\NN^*$ by a factor of order
unity only changes $b$, $y_0$ and $K$ by a similar factor; the
scaling relations of Eqs. (\ref{eq:optimalE}) and
(\ref{eq:optimalA}) are otherwise unchanged.}. For example, in the
case of incoherent Cooper-pair tunneling \cite{Naik06}, $\NN^*$ is
naturally defined as the change of $\NN$ needed take the SET off
resonance. This yields $2 E_C \NN^* \sim \hbar \Gamma$, where $1/
\Gamma$ is the lifetime of the Cooper-pair resonance.

The gate voltage scale $\NN^*$ can be translated into an
energy scale $E^*$ via
\begin{eqnarray}
    E^* = 2 k \left( \frac{E_C \NN^*}{A} \right)^2 .
\end{eqnarray}
This is the approximate energy needed by the oscillator to
completely smear out the negative damping contribution of the SET.
As we will see, $E^*$ is {\it not} the same as the energy $E_0$
discussed above.

Next, we write the SET damping as $\gamma(E) = - \gamma_{max} g(E
/ E^*)$, where $g(0)=1$, and $\gamma_{max} = -\gamma(0)$ is the
maximum of the negative SET damping.  For simplicity, we assume g
is monotonic decreasing, as is the case for an optimally tuned
SSET. The most probable oscillator energy, i.e. the energy at
which a maximum occurs in $w(E)$, is then given by
\begin{eqnarray}
    E_0 & = &
        E^* \cdot g^{-1}\left( \frac{\gamma_0}{ \gamma_{max}} \right) \\
    & = &
        K \cdot Q_0 \hbar \Omega  \cdot
        \left[ \frac{\gamma_0}{ \gamma_{max}} \right]
          g^{-1}\left( \frac{\gamma_0}{ \gamma_{max}} \right) .
          \label{eq:scaling}
\end{eqnarray}
where $Q_0 = \Omega / \gamma_0$ is the intrinsic oscillator
quality factor, and the dimensionless constant $K$ is {\it
independent of the oscillator-SET coupling strength}:
\begin{eqnarray}
    K = \frac{2(E_C \NN^*)^2}{\hbar^2}
     \frac{ d S_n}{d \omega}\bigg|_{\omega=0,\NN=\NN_0}.
\end{eqnarray}
One can easily estimate the parameter $K$ for a given SET
operating point.  For example, we have already shown that $2 E_C
\NN^* \simeq \hbar \Gamma$ for the DJQP process in a SSET.
Further, a simple estimate (which is born out by the full
calculation in Ref.~\onlinecite{Clerk05}) gives $d S_{n} / d
\omega \simeq (\hbar / E_J)^2$ where $E_J$ is the Josephson energy
for the SET (both junctions are assumed equal). We thus have $K
\simeq (\hbar \Gamma/E_J)^2$.  For the experiments of
Ref.~\onlinecite{Naik06}, this yields $K \simeq 6$.

Using Eq.~(\ref{eq:scaling}), we find that we can maximize the
average oscillator energy $E_0$ with respect to the coupling
strength\footnote{The average energy $\langle E \rangle$ is
slightly larger than $E_0$ due to fluctuations; this small
difference plays no role here.}. In that equation, we see that the
coupling strength only enters $E_0$ through the ratio $\gamma_0 /
\gamma_{max}$. Moreover, as $g$ is a monotonic decreasing
function, the same is true of $g^{-1}$.  It thus follows that
$E_0$ must have a maximum as a function of the coupling.
Physically, this is easy to understand.  For too weak a coupling,
the intrinsic damping $\gamma_0$ dominates the effects of the
oscillator, and there is no instability.  For too strong a
coupling, the oscillator does not need much energy to smear out
the effects of the SET-induced damping.

Finally, we define $b$ to be the maximum of the function $y
g^{-1}(y)$, and define $y_0$ to be the value of $y$ at which this
function is maximized.  With our choice of $E^*$, both $b$ and
$y_0$ are of order unity.  We thus have \footnote{In the limit
$Q_0\rightarrow\infty$, we find
$\left[E_0\right]_{max}\rightarrow\infty$ and $A_{opt}\rightarrow
0$. This poses no contradiction with the work of Blanter {\it et
al.} \cite{Blanter05both}, who considered $Q_0\rightarrow\infty$
for finite, {\it fixed} $A$, which leads to finite, but not
optimal, $E_0$.}
\begin{eqnarray}
    \left[ E_0 \right]_{max} & = &
        b \cdot K Q_0 \hbar \Omega ,
            \label{eq:optimalE}\\
    A_{opt} & = & \sqrt{\frac{1}{y_0 \cdot K Q_0}}
        \frac{E_C \NN^*}{\Delta x},
            \label{eq:optimalA}
\end{eqnarray}
where $\Delta x = \sqrt{\hbar / (2 m \Omega)}$ is the zero-point
uncertainty in the oscillator position.

We thus see that for an optimal coupling strength, {\it the
average oscillator energy scales as the intrinsic oscillator
quality factor}. For sufficiently large $Q_0$, one can indeed have
a high mode occupancy in the resonator.  Also note that the
optimal coupling strength scales as $1/\sqrt{Q_0}$.  One can show
along the same lines that in general, $E_0 / Q_0$ only depends on
the coupling and the intrinsic quality factor through the
combination $A \sqrt{Q_0}$.  The validity of these scaling
relations have been tested against numerical solution of
Eq.~(\ref{eq:FP}) for both JQP and DJQP processes
\cite{BennettMSC}.

To further the analogy to a laser, we must also characterize
energy fluctuations in the stationary state. At the optimal
coupling strength, Eq.~(\ref{eq:scaling}) yields the exact result
$\frac{d \gamma}{d E}\big|_{E_0}  = \gamma_0 / E_0$; using this
and assuming that the bath temperature is low enough that the SET
dominates the oscillator diffusion, from Eq.~(\ref{eq:sigma}) we
find
\begin{eqnarray}
    \langle \langle E^2 \rangle \rangle & = &
        \frac{(D(E_0))}{m \gamma_0} \times E_0
        = k_B T^* \cdot E_0.
        \label{eq:variance}
\end{eqnarray}
As in a laser, the variance of $E$ scales as $\sqrt{\langle E
\rangle}$; using Eq.~(\ref{eq:optimalE}), we see that the relative
energy fluctuations become small as $1 / \sqrt{Q_0}$.  The degree
to which the energy fluctuations are super-Poissonian is
determined by the effective temperature $T^*$ defined above. In
practice, $T^*$ will be on the order of the SET effective
temperature at $E=0$, defined by $k_B T_{eff} = D(0) / m
|\gamma(0)|$.  Using the definition of $\gamma$ and $D$ in terms
of $S_n(\omega)$, one can show that in the regime we consider
(where the SET is much faster than the oscillator), $T_{eff}$ and
$T^*$ must both be much larger than $\hbar \Omega$.  Our system is
thus more like a maser than a laser:  the number fluctuations are
far greater than the bound set by purely quantum noise.

{\it Phase fluctuations and linewidth}. Another hallmark of a
laser is its narrow linewidth, which is limited by relatively weak
phase fluctuations in the stationary state\cite{LaxBook}.  It is thus
interesting to ask about phase fluctuations in our system.  A
convenient approach is to work directly with
Eq.~({\ref{eq:Langevin}), and make a rotating-wave approximation,
where one keeps track of the oscillator amplitude and the slowly
varying part of the oscillator phase.  Focusing on fluctuations
about the stationary, large-amplitude lasing state, we start by
writing
\begin{eqnarray}
     \dot{x}(t) - i \tilde{\Omega} x(t) = \left[ \rho_0 + \delta \rho(t) \right]
        e^{-i \tilde{\Omega} t} e^{i \phi(t)},
 \end{eqnarray}
where $\rho_0 = \sqrt{2 E_0 / m}$ is the stationary amplitude of
the oscillator, and $\tilde{\Omega} = \sqrt{(k + \Delta k(E_0)/m}$
is the renormalized frequency of the oscillator, with
\begin{eqnarray}
    \Delta k(E) = 2 \int_0^{2 \pi} \frac{d \theta}{2 \pi}
        \Delta k [x] \sin^2{\theta}.
\end{eqnarray}

The linewidth of the oscillator will be determined by the
fluctuations in the oscillator phase, $\phi(t)$. Linearizing
Eq.~(\ref{eq:Langevin}) in $\delta \rho$ and $d \phi / dt$, and
coarse graining over a timescale long compared to $1 / \Omega$, we
find that the spectral density of this fluctuating phase is
\begin{eqnarray}
    S_{\phi}(\omega) = \frac{2 D_{\phi}}{\omega^2} \left[
        1 + \frac{\alpha^2 }{1 + \left(\omega / \gamma_0\right)^2 }
    \right],
\end{eqnarray}
where the intrinsic phase diffusion constant is given by
\begin{eqnarray}
    D_{\phi} = \frac{1}{4 m E_0}  \int_0^{2 \pi} \frac{d\theta}{2 \pi}
        S_{\delta f}\left[ x  \right] \sin^2{\theta}.
\end{eqnarray}
Here, $\alpha$ characterizes the relative importance of the
coupling between amplitude and phase fluctuations:
\begin{eqnarray}
    \alpha  =
    \frac{ d \Omega / d E}{d \gamma / dE}
            \sqrt{ \frac{D(E_0) / m }{E_0  D_{\phi}} }
    \sim
    \frac{1}{m \Omega} \frac{ d \left(\Delta k\right) / dE}{d \gamma/ dE},
    \label{eq:alpha}
\end{eqnarray}
where the derivatives are evaluated at $E_0$, and we have used the
fact that up to factors of order unity, $D_\phi \sim D(E_0) / (m
E_0)$. $\alpha$ is analogous to Henry's $\alpha$ parameter in
standard laser theory \cite{Henry82}; on a physical level, energy
fluctuations lead to additional frequency fluctuations through the
SET spring constant modulation $\Delta k[x]$.

In a semiclassical picture, the SET-induced damping is due to the small but finite
response time $\tau$ of the SET to the oscillator's motion\cite{Clerk05};
one can write
$m\gamma \sim \left(\Delta k\right) \tau$. From
Eq.~(\ref{eq:alpha}), one then has $\alpha \sim 1/ (\Omega \tau)
\gg 1$, as we are explicitly considering a SET which is much
faster than the oscillator.  This means that unlike a typical
laser, {\it the linewidth of the oscillator in the stationary
state of our ``lasing" NEMS is set by the coupling between
amplitude and phase fluctuations}.  The crucial difference from a
laser is that in the NEMS case, the transfer of energy between the
oscillator and the SET is not resonant, whereas in a laser the
coupling between the mode and the atoms is resonant.  Thus, in a
laser, one obtains a damping force from the atoms even if they are
assumed to respond instantaneously to the mode; in contrast, there
would be no back-action damping in our NEMS in this limit.


Finally, for sufficiently large $E_0$ we find that the oscillator
position noise spectrum exhibits a Lorentzian form centered on
$\omega = \tilde{\Omega}$ with a width  $\gamma_{eff}$ determined
by the zero-frequency limit of $S_{\phi}$.  Up to factors of order
unity, we find that this width can be much narrower than that set
by the intrinsic damping of the oscillator:
\begin{eqnarray}
    \frac{ \gamma_{eff}}{\gamma_0}
    \sim
    \frac{k_B T^*}{4E_0} \alpha^2
    \propto \frac{1}{K Q_0}.
\end{eqnarray}

{\it Critical slowing down and shot noise.}
\begin{figure}
\center{\includegraphics[width=8cm]{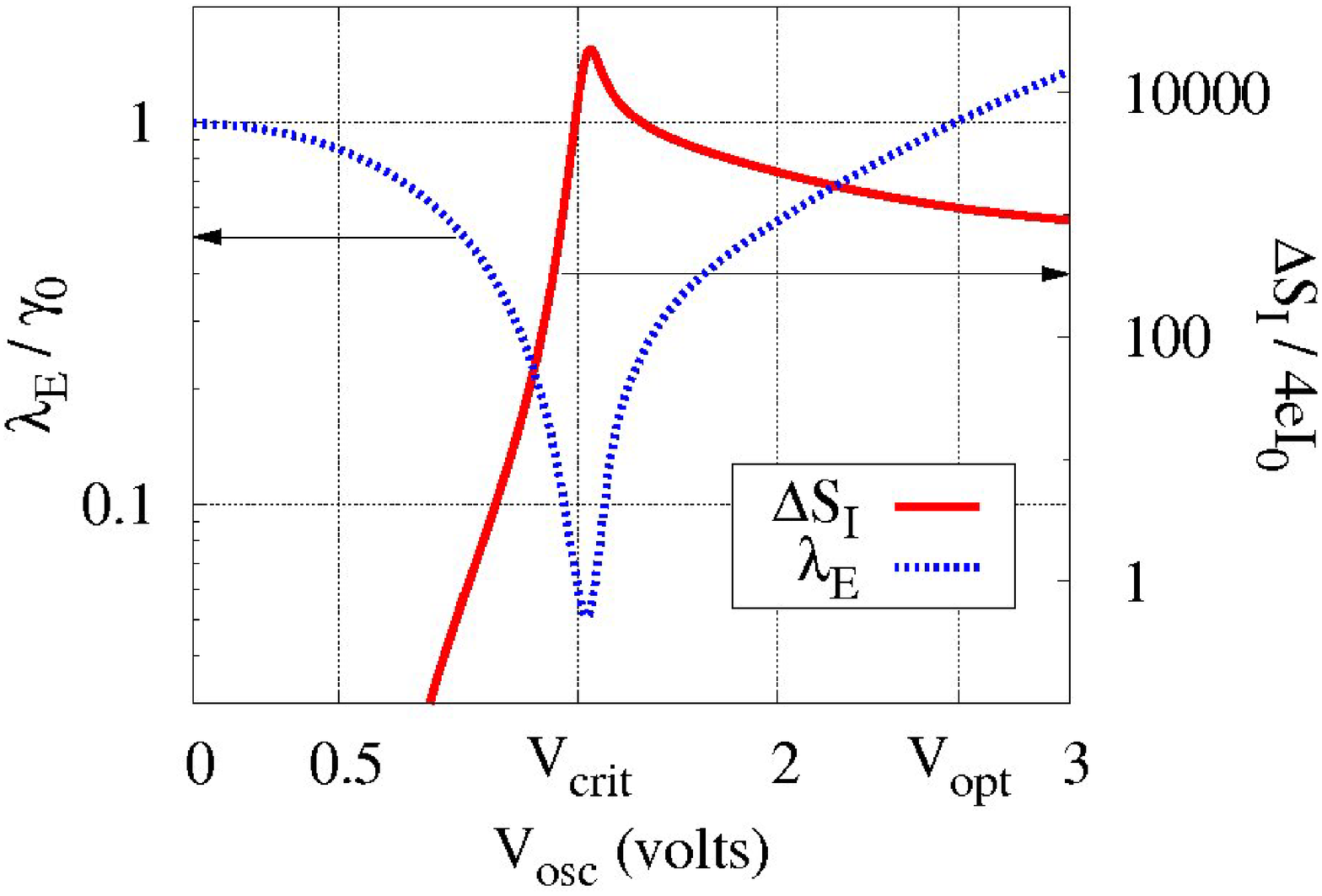}}
\caption{Signatures of critical slowing down for the DJQP process
in a SSET.  Dotted line (left vertical axis): energy relaxation
rate $\lambda_E$ versus coupling voltage $V_{osc}$. The minimum near
$V_{crit}$ corresponds to critical slowing down, and one finds
$\lambda_E = \gamma_0$ at the optimal coupling voltage, $V_{opt}$.
Solid line (right vertical axis): zero-frequency current noise
enhancement $\Delta S_I$ versus $V_{osc}$, scaled by the
current noise of the uncoupled SSET. 
We have used device
parameters from Ref. \onlinecite{Naik06}; for the SSET, these are
$E_C = 175\ \mu\text{eV}$, $\Delta = 192\ \mu\text{eV}$, $g =
0.5$, and $E_J = 15\ \mu\text{eV}$. For the oscillator: $\Omega =
22$ MHz, $k = 10$ N/m and $Q_0 =$ 120 000.  We have also used
$T_0 = 20$ mK and $A = 1.05 \times 10^{-13}$ N when
$V_{osc}=1$ volt. } \label{fig:noise} \vspace{-0.5 cm}
\end{figure}
As was discussed in Ref.~\onlinecite{Blanter05both}, the
instability physics discussed here will manifest itself directly
in the shot noise of the SET.  In the stationary state, the
oscillator's motion is large enough that the resulting
oscillations in the SET gate voltage $\NN$ strongly modulate the
current. One consequence is that the slow energy dynamics of the
SET will lead to long-time correlations of tunnel events in the
SET, and thus a large enhancement of the low-frequency current
noise.   The magnitude of this enhancement is expected to be
inversely proportional to the relaxation rate of energy
fluctuations in the oscillator, $\lambda_E$: this represents the time over which 
tunnel events are correlated.  At the optimal
coupling, a linearized treatment of fluctuations (as above) shows
that $\lambda_E =  \gamma_0$. Moreover, as a function of coupling
strength, $\lambda_E$ exhibits critical slowing down: it reaches a
minimum as the coupling strength is tuned through the bifurcation
between small-amplitude and large-amplitude states of the
oscillator. Without  noise, 
this occurs at the coupling voltage $V_{crit}$ for
which $\gamma(E=0)+\gamma_0=0$; fluctuations push the transition
to a slightly higher coupling voltage \cite{LaxBook}.  We thus expect a
maximum shot noise enhancement near $V_{osc}= V_{crit}$.


We have numerically calculated the expected shot noise enhancement
for a NEMS system operated near the DJQP resonance, using
parameters from Ref. \onlinecite{Naik06}.  This can be calculated
in a straightforward manner from the energy diffusion equation Eq.
(\ref{eq:FP}) \cite{Blanter05both}.  Fig. \ref{fig:noise} shows a
clear maximum in the shot noise as a function of the coupling
voltage, which corresponds to the critical slowing down.  We
stress that the maximum does not occur at the optimal coupling
voltage $V_{opt}$, but rather near the voltage $V_{crit}$
corresponding to the bifurcation.

In conclusion, we have shown how negative damping instabilities in
SET NEMS scale with system parameters, and have investigated the
amplitude and phase fluctuations in the stationary state.  This
work was supported by NSERC and FQRNT.

\bibliographystyle{apsrev}
\bibliography{AashRefsMod}

\begin{thebibliography}{18}
\expandafter\ifx\csname natexlab\endcsname\relax\def\natexlab#1{#1}\fi
\expandafter\ifx\csname bibnamefont\endcsname\relax
  \def\bibnamefont#1{#1}\fi
\expandafter\ifx\csname bibfnamefont\endcsname\relax
  \def\bibfnamefont#1{#1}\fi
\expandafter\ifx\csname citenamefont\endcsname\relax
  \def\citenamefont#1{#1}\fi
\expandafter\ifx\csname url\endcsname\relax
  \def\url#1{\texttt{#1}}\fi
\expandafter\ifx\csname urlprefix\endcsname\relax\def\urlprefix{URL }\fi
\providecommand{\bibinfo}[2]{#2}
\providecommand{\eprint}[2][]{\url{#2}}

\bibitem[{\citenamefont{Blencowe and Wybourne}(2000)}]{Blencowe00}
\bibinfo{author}{\bibfnamefont{M.~P.} \bibnamefont{Blencowe}} \bibnamefont{and}
  \bibinfo{author}{\bibfnamefont{M.~N.} \bibnamefont{Wybourne}},
  \bibinfo{journal}{Appl. Phys. Lett.} \textbf{\bibinfo{volume}{77}},
  \bibinfo{pages}{3845} (\bibinfo{year}{2000}).

\bibitem[{\citenamefont{Knobel and Cleland}(2003)}]{Knobel03}
\bibinfo{author}{\bibfnamefont{R.~G.} \bibnamefont{Knobel}} \bibnamefont{and}
  \bibinfo{author}{\bibfnamefont{A.~N.} \bibnamefont{Cleland}},
  \bibinfo{journal}{Nature (London)} \textbf{\bibinfo{volume}{424}},
  \bibinfo{pages}{291} (\bibinfo{year}{2003}).

\bibitem[{\citenamefont{LaHaye et~al.}(2004)\citenamefont{LaHaye, Buu,
  Camarota, and Schwab}}]{LaHaye04}
\bibinfo{author}{\bibfnamefont{M.~D.} \bibnamefont{LaHaye}},
  \bibinfo{author}{\bibfnamefont{O.}~\bibnamefont{Buu}},
  \bibinfo{author}{\bibfnamefont{B.}~\bibnamefont{Camarota}}, \bibnamefont{and}
  \bibinfo{author}{\bibfnamefont{K.~C.} \bibnamefont{Schwab}},
  \bibinfo{journal}{Science} \textbf{\bibinfo{volume}{304}},
  \bibinfo{pages}{74} (\bibinfo{year}{2004}).

\bibitem[{\citenamefont{Naik et~al.}(2006)\citenamefont{Naik, Buu, LaHaye,
  Armour, Clerk, Blencowe, and Schwab}}]{Naik06}
\bibinfo{author}{\bibfnamefont{A.}~\bibnamefont{Naik}},
  \bibinfo{author}{\bibfnamefont{O.}~\bibnamefont{Buu}},
  \bibinfo{author}{\bibfnamefont{M.~D.} \bibnamefont{LaHaye}},
  \bibinfo{author}{\bibfnamefont{A.~D.} \bibnamefont{Armour}},
  \bibinfo{author}{\bibfnamefont{A.~A.} \bibnamefont{Clerk}},
  \bibinfo{author}{\bibfnamefont{M.~P.} \bibnamefont{Blencowe}},
  \bibnamefont{and} \bibinfo{author}{\bibfnamefont{K.~C.}
  \bibnamefont{Schwab}}, \bibinfo{journal}{Nature (London)}
  \textbf{\bibinfo{volume}{443}}, \bibinfo{pages}{193} (\bibinfo{year}{2006}).

\bibitem[{\citenamefont{Blencowe}(2004)}]{Blencowe04}
\bibinfo{author}{\bibfnamefont{M.~P.} \bibnamefont{Blencowe}},
  \bibinfo{journal}{Phys. Rep.} \textbf{\bibinfo{volume}{395}},
  \bibinfo{pages}{159} (\bibinfo{year}{2004}).

\bibitem[{\citenamefont{Mozyrsky and Martin}(2002)}]{Mozyrsky02}
\bibinfo{author}{\bibfnamefont{D.}~\bibnamefont{Mozyrsky}} \bibnamefont{and}
  \bibinfo{author}{\bibfnamefont{I.}~\bibnamefont{Martin}},
  \bibinfo{journal}{Phys. Rev. Lett.} \textbf{\bibinfo{volume}{89}},
  \bibinfo{pages}{018301} (\bibinfo{year}{2002}).

\bibitem[{\citenamefont{Armour et~al.}(2004)\citenamefont{Armour, Blencowe, and
  Zhang}}]{Armour04}
\bibinfo{author}{\bibfnamefont{A.~D.} \bibnamefont{Armour}},
  \bibinfo{author}{\bibfnamefont{M.~P.} \bibnamefont{Blencowe}},
  \bibnamefont{and} \bibinfo{author}{\bibfnamefont{Y.}~\bibnamefont{Zhang}},
  \bibinfo{journal}{Phys. Rev. B} \textbf{\bibinfo{volume}{69}},
  \bibinfo{pages}{125313} (\bibinfo{year}{2004}).

\bibitem[{\citenamefont{Clerk}(2004)}]{Clerk04c}
\bibinfo{author}{\bibfnamefont{A.~A.} \bibnamefont{Clerk}},
  \bibinfo{journal}{Phys. Rev. B} \textbf{\bibinfo{volume}{70}},
  \bibinfo{pages}{245306} (\bibinfo{year}{2004}).

\bibitem[{Bla()}]{Blanter05both}
\bibinfo{note}{Ya. M. Blanter, O. Usmani, and Yu. V. Nazarov, Phys. Rev. Lett.
  {\bf 93}, 136802 (2004); ibid., Phys. Rev. Lett. (erratum) {\bf 94}, 049904
  (2005); O. Usmani, Ya. M. Blanter, and Yu. V. Nazarov, cond-mat/060317}

\bibitem[{\citenamefont{Blencowe et~al.}(2005)\citenamefont{Blencowe, Imbers,
  and Armour}}]{Blencowe05}
\bibinfo{author}{\bibfnamefont{M.~P.} \bibnamefont{Blencowe}},
  \bibinfo{author}{\bibfnamefont{J.}~\bibnamefont{Imbers}}, \bibnamefont{and}
  \bibinfo{author}{\bibfnamefont{A.~D.} \bibnamefont{Armour}},
  \bibinfo{journal}{New J. Phys.} \textbf{\bibinfo{volume}{7}},
  \bibinfo{pages}{236} (\bibinfo{year}{2005}).

\bibitem[{\citenamefont{Clerk and Bennett}(2005)}]{Clerk05}
\bibinfo{author}{\bibfnamefont{A.~A.} \bibnamefont{Clerk}} \bibnamefont{and}
  \bibinfo{author}{\bibfnamefont{S.}~\bibnamefont{Bennett}},
  \bibinfo{journal}{New J. Phys.} \textbf{\bibinfo{volume}{7}},
  \bibinfo{pages}{238} (\bibinfo{year}{2005}).

\bibitem[{\citenamefont{Rodrigues and Armour}(2006)}]{Armour06}
\bibinfo{author}{\bibfnamefont{D.~A.} \bibnamefont{Rodrigues}}
  \bibnamefont{and} \bibinfo{author}{\bibfnamefont{A.~D.}
  \bibnamefont{Armour}}, \bibinfo{journal}{cond-mat/0608166}
  (\bibinfo{year}{2006}).

\bibitem[{\citenamefont{Bargatin and Roukes}(2003)}]{Bargatin03}
\bibinfo{author}{\bibfnamefont{I.}~\bibnamefont{Bargatin}} \bibnamefont{and}
  \bibinfo{author}{\bibfnamefont{M.~L.} \bibnamefont{Roukes}},
  \bibinfo{journal}{Phys. Rev. Lett.} \textbf{\bibinfo{volume}{91}},
  \bibinfo{pages}{138302} (\bibinfo{year}{2003}).

\bibitem[{\citenamefont{Haken}(1975)}]{Haken75}
\bibinfo{author}{\bibfnamefont{H.}~\bibnamefont{Haken}}, \bibinfo{journal}{Rev.
  Mod. Phys.} \textbf{\bibinfo{volume}{47}}, \bibinfo{pages}{67}
  (\bibinfo{year}{1975}).

\bibitem[{\citenamefont{Dykman and Krivoglaz}(1984)}]{Dykman84}
\bibinfo{author}{\bibfnamefont{M.~I.} \bibnamefont{Dykman}} \bibnamefont{and}
  \bibinfo{author}{\bibfnamefont{M.~V.} \bibnamefont{Krivoglaz}}, in
  \emph{\bibinfo{booktitle}{Soviet Physics Reviews}}, edited by
  \bibinfo{editor}{\bibfnamefont{I.~M.} \bibnamefont{Khalatnikov}}
  (\bibinfo{publisher}{Harwood Academic}, \bibinfo{year}{1984}),
  vol.~\bibinfo{volume}{5}, pp. \bibinfo{pages}{265--441}.

\bibitem[{\citenamefont{Bennett}(2006)}]{BennettMSC}
\bibinfo{author}{\bibfnamefont{S.}~\bibnamefont{Bennett}}, Master's thesis,
  \bibinfo{school}{McGill University} (\bibinfo{year}{2006}).

\bibitem[{\citenamefont{Lax}(1966)}]{LaxBook}
\bibinfo{author}{\bibfnamefont{M.}~\bibnamefont{Lax}}, in
  \emph{\bibinfo{booktitle}{Brandeis Summer Institute of Theoretical Physics
  Lectures}} (\bibinfo{publisher}{Gordon and Breach}, \bibinfo{year}{1966}).

\bibitem[{\citenamefont{Henry}(1982)}]{Henry82}
\bibinfo{author}{\bibfnamefont{C.~H.} \bibnamefont{Henry}},
  \bibinfo{journal}{IEEE J. Quantum Electron.} \textbf{\bibinfo{volume}{18}},
  \bibinfo{pages}{259} (\bibinfo{year}{1982}).

\end{thebibliography}

\end{document}